# LOGICAL ANALYSIS AND CONTRADICTION DETECTION IN HIGH-LEVEL REQUIREMENTS DURING THE REVİEW PROCESS USING SAT-SOLVER

Simge Yatkın[1,2] and Tolga Ovatman[1]

[1]Faculty of Computer and Informatics Engineering, Istanbul Technical University, Istanbul, Turkey
[2]MGEO Test and Verification Directorate, ASELSAN Inc., Ankara, Turkey

*ABSTRACT*

*DO-178C stands out as a guiding standard for aviation system development processes. This standard not only mandates ensuring the consistency of requirements in the software verification process but also recognizes it as a mandatory element. The main objective of this study is to introduce a method for analyzing and identifying inconsistencies between high-level requirements using information obtained from a data dictionary. This method aims to transform high-level requirements into logical expressions and then thoroughly examine them using a SAT Solver to detect inconsistencies. While methods focused on identifying inconsistencies among requirements often appear in the literature, this study presents a novel approach to detect contradictions between non-natural language, systematically structured, and language-independent requirements. The goal of this approach is to significantly reduce the review time of high-level requirements in the software verification process. Evaluations indicate that the use of this method results in substantial time savings in the inconsistency detection process.*

## 1. INTRODUCTION

DO-178C stands as an obligatory standard for aviation certification authorities, including the Federal Aviation Administration (FAA), the European Union Aviation Safety Agency (EASA), and various other certification entities. In today's aviation landscape, this standard provides comprehensive guidance for the software-based certification processes of commercial aviation systems, considering the complexity of modern aviation systems. Crucial for the safety and performance of the aviation industry, DO-178C requires a thorough examination of all process outputs to determine the accuracy of the software and identify potential errors.

Furthermore, the standard emphasizes the consistency of specific high-level requirements at various stages of the software development process. In particular, ensuring the consistency of these high-level requirements, as detailed in section 6.3.1 of the DO-178C document [1], is a critical element for the successful progression of the certification process. This requirement aims to secure the compliance of the software with civil aviation standards, ensuring adherence to industry norms.

In complex software systems, requirements can conflict over time for various reasons. Failure to address these conflicts in the early stages can lead to issues and errors in the software development process. Therefore, the early detection of inconsistencies is crucial for ensuring a smooth software development process. Consistency among requirements requires that two or more requirements do not contradict each other[2]. This becomes particularly challenging and time-consuming in large-scale software projects, such as avionics, where there are often more than ten modules, each containing hundreds of requirements. Given the complexity of requirements, the workload increases, and reviewers may be more prone to making errors. The identification of these conflicts during the writing of tests and the testing of real systems can delay standardized processes. This underscores the importance of early and meticulous attention to requirement consistency to prevent potential setbacks in the development timeline.

This study presents a method aiming to automate the analysis of inconsistencies among requirements by utilizing data specified in the data dictionary during the review process of high-level requirements. The proposed methodology integrates with the Dynamic Object-Oriented Requirements System (DOORS) to transform each high-level requirement in the High-Level Requirements (HLR) document into logical expressions. The requirements extracted from the relevant module within DOORS are converted into logical expressions for various programming languages and data formats using ANTLR4 (ANother Tool for Language Recognition), a versatile parser generator. These logical expressions are then subjected to analysis using a SAT Solver.

The primary objective of this approach is to reduce the time spent on reviewing high-level requirements in the software verification process and minimize potential reviewer errors. The implementation of this method has the potential to significantly enhance the efficiency and accuracy of the software development process.

## 2. RELATED WORKS

In recent times, there has been a significant increase in interest in the analysis of conflicts in software development processes. Both researchers and practitioners acknowledge the critical role of early detection and resolution of conflicts in the development process in enhancing software quality and minimizing costly revisions. Various methods are proposed in studies focusing on this important topic to detect and resolve conflicts among requirements[3].

A method proposed by Egyed and Grunbacher[4] utilizes automated traceability techniques to eliminate conflicts. This approach automatically identifies conflicts based on features among requirements and determines traceability dependencies. Another system developed by Kim and others[5] uses the RECOMA tool to detect conflicts based on natural language requirement segments. This system employs both syntactic and semantic methods for conflict detection. Moser et al. [6],[7] present an automatic semantic-based conflict detection approach that associates natural language requirements with semantic concepts and defines conflicts using logical expressions. Urbieta and the team [8],[9] propose a model-based approach aiming to detect requirement conflicts in the early stages of software development for web applications. Chentouf's [10] method for resolving OAM\&P requirement conflicts involves representing requirements using EBNF and determining conflicts based on inference rules. Comparison between existing (automatic method) and the method we present can be observed in Table 1.

Software requirements are defined as a collection of English expressions summarizing the intended functionality of the software. Zowghi [11] has successfully demonstrated that these English expressions can be translated into equivalent logical statements. Automatic theorem provers [12], [13] can be utilized for the purpose of detecting logical inconsistencies. However, in some companies, High-Level Requirements (HLR) documents may include non-English expressions, as they rely on data in the data dictionary, leading to diversity, including abbreviations and proper names. Various studies offer different approaches for the detection and management of requirement conflicts. Nevertheless, there is limited research in the existing literature on conflict analysis methods based on HLRs and related data dictionaries. This study aims to contribute by providing a different approach to the review process in areas with similar software requirements.

Table 1. Comparison between existing (automatic method) and our presented studies.

| Reference | Approach to Analyzing Conflicts | Method for Identifying Conflicts | Requirements Category | Representation of Requirements |
|---|---|---|---|---|
| [4] | Traceability approach | Automatic | Functional & Nonfunctional | - |
| [5] | Natural Language Partitioning of Requirements | Automatic | Functional | Formalization |
| [6], [7] | Semantic based approach | Automatic | Functional | Ontology |
| [8], [9] | Utilization of NDT Meta Model in Graphical Techniques | Automatic | Functional | Formalization(DSL) Structure model(NDT requirement meta model) |
| [10] | Validation rules | Automatic | Functional | Formalization |
| This study | Satisfiable approach | Automatic | Functional & Nonfunctional | Logical Expression |

## 3. BACKGROUND

### 3.1. IBM Rational DOORS and Data Dictionary

High-Level Requirements (HLR) define in detail the expected behavior of all software installed on the target computer, independent of software architecture. These comprehensive requirements encompass a broad dataset, particularly considering the complex structure of avionic systems [14]. This dataset is organized by a data dictionary, which essentially presents data element names and their definitions in a tabular format [15]. The dictionary includes the values or value ranges that the data within the requirements can take.

Some companies prefer to create a Data Definition Table (DDT) that aligns with the purpose of the data dictionary. In Table 2, it can be observed, for example, how the values corresponding to the data names are presented in the Range/Unit column. DDTs are found on the platform where software requirements are stored. In our context, we manage our requirements within IBM Rational DOORS. IBM Rational DOORS is a leading requirement management solution that provides a collaborative environment for defining, capturing, and managing requirements throughout the development lifecycle~cite [19].

Table 2. Data definition table example.

| Data | Range / Unit |
|---|---|
| DCU_Type | DCU_1, DCU_ 2 |
| BIT_Status | UNKNOWN, PBIT_RESULT, IBIT_RESULT |
| ABC_Status | PASS, FAIL |
| SjRequestCond | TRUE, FALSE |

This integrated platform facilitates the systematic tracking and updating of requirements, enabling more effective project management. Specifically designed to alleviate the complexity in the software development process and optimize the requirement management process, IBM Rational DOORS stands out as a valuable tool that efficiently manages detailed requirements.

### 3.2. DOORS Extension Language (DXL)

DXL is a specialized scripting language designed for use with IBM Rational DOORS. It allows users to automate various tasks, customize the DOORS environment, and extract information from DOORS modules programmatically. DXL scripts are instrumental in streamlining workflows and enhancing the overall efficiency of the requirements management process. The integration process involves exporting requirements from DOORS using DXL script.

### 3.3. Requirement Format

In conducting this study, we based our analysis on requirements presented in two different formats. The first is a requirements format commonly used in avionics software. The second format is a customized version of the Gherkin language [16] shown in Figure 1. Keywords such as Given, Then, And and Or are used. *Conditions* are specified from "Given" to "Then", after "Then" the *operations* are specified.

```
1    ID:Gherkin_1
2
3    Given MMM is SJ
4    And MOS_S is NOT DGFT
5    And SjRequestCond is NOT TRUE
6    And MOS_Status is NONE
7    Then SET MMM to NAV
```

Figure 1. A requirement written in customized version of the Gherkin

To systematically analyze the contradictions between requirements, we started the process by creating a dataset. Within these datasets, requirements were rigorously parsed based on the grammars associated with their respective formats. This parsing process was facilitated using ANTLR4 (ANother Tool for Language Recognition), a versatile parser known for its ability to create parsers for various programming languages and data formats [17].

## 3.4. Jenkins and Reporting

Jenkins is an open source automation server that automates the build, test and deployment processes of software development, facilitating continuous integration and supporting continuous delivery. We automate our work with Jenkins and after each run, the results of the conflict analysis are reported in HTML format. In this way, continuous analysis work can be maintained between changing requirements. With the report generated, employees responsible for the review process can review the report and make requests for necessary changes.

## 4. PROPOSED METHOD

The overall objective of the proposed method is to analyze logical contradictions between requirements with common operations using SAT-Solver. In order to perform this analysis, we propose a method (also see Figure 2) consisting of the three phases.

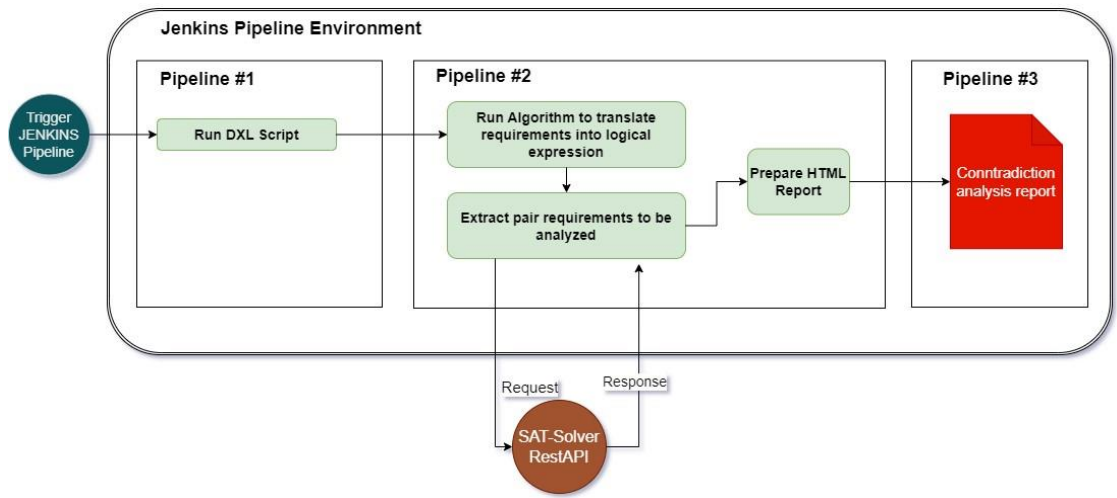

Figure 2. Requirements conflict analysis system

## 4.1. Converting Requirements to Logical Expressions

With ANTLR4, a grammar for requirements formats is written, as shown in Figure 3 and requirements parsing is performed. With the grammar we have written, the requirements received from DOORS with the DXL script (See Figure 2, Pipeline #1 box) are converted into logical expressions as a result of the parsing algorithm (See Figure 2, Pipeline #2 box).

Conditions and operations in the requirements are meticulously defined and assigned unique characters for reference and manipulation. Boolean conditions were cleverly represented, reducing complex sentences to single characters and thus simplifying our analysis. Conditions without Boolean attributes were approached differently through the application of negation.

In the requirement example given in Figure 1, we obtain the results in the tables from conditions (see Table 3) and operations (see Table 4). Once all conditions and operations have been assigned in this way, the requirements are translated into logical expressions. As a result of these inferences, the logical expression equivalent of the requirement shown in Figure 1 is as follows:

*Logical Expression of Gherkin1 requirement: And(A, Not(B), Not(C), D) => X*

```
1  grammar Gherkin;
2  start: (reqID condition operation)+;
3  condition: GIVEN (req | nested) (and | or)*;
4  operation: WHEN (req | nested) (and | or)*;
5  and: AND (req | nested);
6  or: OR (req | nested);
7  reqID: 'ID:' WORD NEWLINE;
8  req: (boolReq | otherReq) ;
9  nested: '(' nestedReq ')' NEWLINE*;
10 nestedReq: req (('And ' |'Or ') req)+;
11 boolReq: (CHAR* WORD+ CHAR*)+ BOOLEAN NEWLINE*;
12 otherReq: (CHAR* WORD+ CHAR*)+ NEWLINE*;
13 GIVEN: 'Given ';
14 WHEN: 'When ';
15 PHOPEN: '(';
16 PHCLOSE: ')';
17 AND: 'And ';
18 OR: 'Or ';
19
```

Figure 3. An overview of the grammar of the customized Gherkin format

Table 3. Illustration of the conditions derived from the example requirement shown in Figure 1.

| Conditions | Symbols |
| --- | --- |
| MMM is SJ | A |
| MMM is NOT SJ | Not(A) |
| MOS_S is DGFT | B |
| MOS_S is NOT DGFT | Not(B) |
| SjRequestCond is TRUE | C |
| SjRequestCond is NOT TRUE | Not(C) |
| SjRequestCond is FALSE | Not(C) |
| SjRequestCond is NOT FALSE | C |
| MOS_Status is NONE | D |
| MOS_Status is NOT NONE | Not(D) |

Table 4. Illustration of the operations derived from the example requirement shown in Figure 1.

| Operations | Symbols |
| --- | --- |
| MMM to NAV | X |
| MMM to NOT NAV | Not(X) |

Furthermore, with the approach we present, we are able to analyze contradictions in nested conditions. Nested conditions refer to requirements that have "Or" inside conditions with "And" or "And" inside conditions with "Or". Looking at the requirements given in Figure 4, the conflict analysis is performed between two requirements, called Req1 and Req2, because they have the same operations. The operations in the given requirements are the same. Even if only one of them was the same, the same analysis would still be performed. As a result of the algorithm we developed, the logical expressions corresponding to Req1 and Req2 requirements are as follows:

*Logical Expression of Req1 requirement: Or(A, And(Not (B), C)) => And(X, Y)*
*Logical Expression of Req2 requirement: Or( Not (A), And(B, Not (C) ) => And(X, Y)*

```
1    Req1:
2    Given Op1Cond is TRUE
3    Or (MPoint is NOT OA And Op2Cond is TRUE)
4    Then TPoint to TKPoint
5    And SET MODE to LockMode
6
7    Req2 :
8    Given Op1Cond is FALSE
9    Or (MPoint is OA And Op2Cond is FALSE)
10   Then TPoint to TKPoint
11   And SET MODE to LockMode
```

Figure 4. Example HLR requirements

The requirements given in Figure 4 are extracted from the DOORS module prepared for the study and represent conflicting requirements. These requirements that share the same operations are analyzed using SAT-Solver. The SAT-Solver is given an expression of the following form:

*And(Or(A, And(Not(B), C)), Or(Not(A), And(B, Not(C))))*

Since no solution is available, the SAT Solver produces an output indicating the existence of a contradiction.

In addition, our method involves discovering contradictions within HLR requirements, the "Hypothetical Syllogism" rule for nuanced contradictions. The general form of the hypothetical syllogism is as follows:

*If P, then Q.*
*If Q, then R.*
----------------------------------------
*Therefore, if P, then R.*

An operation in one requirement may be a condition of another requirement. This can lead to contradictions, and manually finding these contradictions can be challenging and time-consuming.

### 4.2. SAT-Solver Rest API

Contradiction analysis is conducted among requirement pairs with shared operations. The request, containing requirement pairs translated into logical expressions, is sent to the SAT-Solver Rest API utilizing the Sympy [18] library. The operation within this API involves subjecting the condition clauses of requirement pairs to an AND operation. If there is no solution set in this process, it returns a response indicating the presence of a contradiction.

For example, the requirements P → Q and R → Q have the same operation. These pairs are passed as input to the SAT-Solver. To analyze these pairs, we apply a logical AND operation to the condition parts. If this AND operation results in the value Zero (0), i.e. no solution set can be found using the SAT-Solver, this indicates the existence of a contradiction.

### 4.3. Contradiction Analysis Report

At the end of this study, reporting holds a significant place. Our work, which is connected to Jenkins automation (see Figure 5), can be triggered automatically whenever users request or after the update of requirement documents.

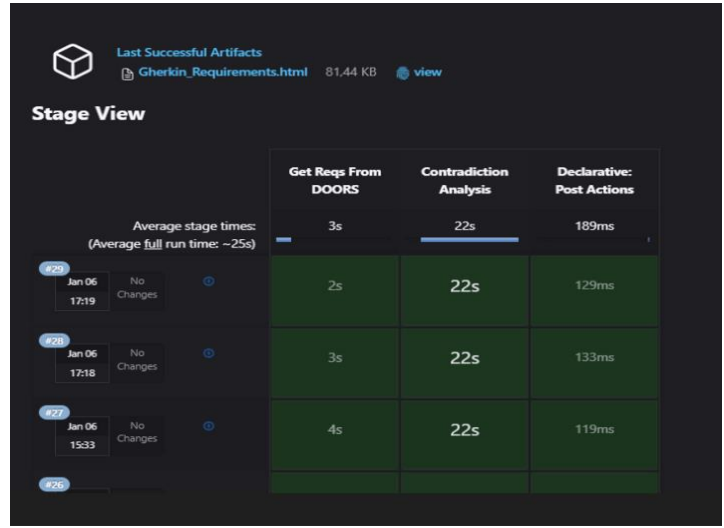

Figure 5. Stage view of Jenkins pipeline for requirement contradiction analysis system

The results of the analyses conducted with the parser algorithm implemented in our study and the SAT-Solver used are presented in an HTML document. This document includes the analysis results of all compared requirements. Additionally, if there are contradictions arising from the application of the "Hypothetical Syllogism" rule, it is documented how these contradictions are created based on this rule.

Figure 6. Contradiction Analysis Report

The presented method has certain limitations. Firstly, if there are requirements written based on a standard, the grammar of this standard should be created using ANTLR4. Subsequently, additional implementations should be made for logical transformations. Moreover, since our method does not incorporate any natural language processing technique, the transformation of logical expressions in requirement sets written in natural language must be done manually. The overall effectiveness and applicability of the method should be carefully assessed in various software development contexts and scenarios. To overcome this limitation, we plan to integrate NLP into our method in future developments. The performance of the method may be subject to certain constraints when dealing with complex requirement sets in large-scale software projects. Another limitation is that parsing times for very large logical formulas can be lengthy. However, when comparing our method with the manual conflict detection approach, it can still be argued that time is saved.

## 5. Evaluatİon

A dataset of 25 requirements covering avionics HLR was prepared. These requirements went through multiple revisions. Initially, the requirements were not contradictory; however, for the purpose of the analysis, the conditions and operations in the dataset were modified. The dataset included simple requirements, slightly more complex requirements with nested conditions, and requirements where operations served as conditions for other requirements.

Ten participants were given a maximum of half an hour to identify logical contradictions in the HLR requirements. They were asked to note down the contradictions they found and to record the time taken to identify each contradiction. Participants' experience with HLR requirements ranged from 2 to 5 years.

A notable finding was that participants were unable to identify contradictions when operations served as a condition for other requirements. With the developed method, 25 requirements were internally compared within 25 seconds, revealing a total of six contradictions. One contradiction was identified using the Hypothetical Comparison rule and interestingly, none of the participants found this contradiction. Furthermore, 4 correct contradictions were detected in an average of 21 minutes. Considering large-scale systems with thousands of requirements, uncovering contradictions between requirements would be a time-consuming process.

## 6. Conclusion

Consequently, the software verification process, especially for commercial aviation systems subject to certification standards such as DO-178C, requires a thorough review of the high-level requirements (HLRs). The presented method aims to streamline the process and reduce the workload associated with the review of complex requirements by using logical expressions and SAT-Solver analysis to detect contradictions between HLRs. The results obtained demonstrate the effectiveness of this approach in identifying contradictions between requirements. This study introduces a methodology that has the potential to increase consistency in software requirements and consequently contribute to improvements in software development processes.

Future studies will have a more comprehensive evaluation process. 4 different data sets with 10, 20, 30 and 40 different requirements will be prepared. Depending on the time, employees with different levels of expertise will be asked to analyze how many contradictions they found in 4 different data sets in how much time. In this way, the difference in time spent between manual and automated approach will be more clearly demonstrated.